\documentclass[12pt]{article}
\usepackage{amssymb,amsmath,epsfig}

\begin{document}

\title{\bf Thermodynamics in Closed Universe with Entropy Corrections}
\author{M. Sharif \thanks {msharif.math@pu.edu.pk} and Abdul Jawad\thanks {jawadab181@yahoo.com}\\
Department of Mathematics, University of the Punjab,\\
Quaid-e-Azam Campus, Lahore-54590, Pakistan.}

\date{}

\maketitle
\begin{abstract}
We discuss the generalized second law of thermodynamics in three
different systems by taking quantum corrections (logarithmic and
power law) to cosmological horizon entropy as well as black hole
entropy. Firstly, we consider phantom energy accretion onto the
Schwarzschild black hole in the closed FRW universe and investigate
validity of the generalized second law of thermodynamics on the
apparent and event horizons. In another scenario, we evaluate the
critical mass of the Schwarzschild black hole with upper and lower
bounds under accretion process due to phantom-like modified
generalized chaplygin gas. It is found that the generalized second
law of thermodynamics is respected within these bounds and black
hole cannot accrete outside them. Finally, we explore this law for a
closed universe filled with interacting $n$-components of fluid (in
thermal equilibrium case) and with non-interacting dark matter and
dark energy components (in thermal non-equilibrium case) on the
apparent and event horizons and find conditions for its validity.
\end{abstract}

\section{Introduction}

Recent developments$^{{1-4})}$ provide indication that our universe
is expanding and has entered in the accelerating phase. In this
respect, the unknown force called dark energy (DE) plays a crucial
role. The value of the equation of state (EoS) parameter
differentiates three different phases of DE according to the
expansion rate of the universe such as quintessence
$(-1<\omega<-\frac{1}{3})$, vacuum ($\omega=-1$) and phantom
($\omega<-1$). Nowadays, phantom cosmology has been taken into
account with great deal of interest. According to it, the energy
density of the phantom energy will turn out to be infinite in a
finite time. This scenario corresponds to a type of future
singularity named as big rip.

It is predicted that phantom energy with large negative pressure
would unstable and disperse all the gravitationally bounded objects
near big rip. This is also confirmed through numerical analysis on
the milky way galaxy and solar system$^{{5})}$. Babichev et
al.$^{{6})}$ modeled the scenario of accretion of phantom energy
onto the Schwarzschild black hole (BH) and observed that BH mass
would gradually decrease and disappear at the end near the big rip.
Yurov et al.$^{{7})}$ found a big hole phenomenon (where BH grows up
rapidly and gains infinitely large size before big rip) for a
specific choice of scale factor. Later, some people$^{{8,9})}$
confirmed the results of Babichev et al.$^{{6})}$ about the
diminishing of mass of BH due to phantom energy accretion for
different BHs, e.g., Reissner-N$\ddot{o}$rdstrom$^{{10})}$,
Kerr-Newmann$^{{9,11})}$, Schwarzschild-de Sitter$^{{9})}$ etc.
Jamil$^{{12})}$ studied the accretion of phantom-like chaplygin gas
models such as modified variable chaplygin gas and viscous
generalized chaplygin gas onto the Schwarzschild BH. He showed that
accretion process decreases the mass of BH in the violation of null
energy condition. Bhadra and Debnath$^{{13})}$ have extended this
work for the Schwarzschild and Kerr-Newmann BHs by using new
variable modified chaplygin gas and generalized cosmic chaplygin
gas.

In addition, the loss of BH masses due to accreting phantom fluid
causes decrease in the entropies since the area of BH is directly
proportional to its entropy. In this scenario, the status of
thermodynamical quantities like entropy as well as temperature and
correspondingly the validity of the generalized second law of
thermodynamics (GSLT) has become an interesting subject. Izquierdo
and Pav$\acute{o}$n have discussed the GSLT for phantom dominated
universe in the absence$^{{14})}$ and presence of the Schwarzschild
BH$^{{15})}$. Further, Sadjadi found different conditions for the
validity of GSLT for the phantom dominated universe enclosed by the
future event horizon in the absence$^{{16})}$ and presence of the
Schwarzschild BH$^{{17})}$. In the meantime, Pacheco and
Horvath$^{{18})}$ investigated GSLT of a system containing the
phantom fluid and the Schwarzschild BH by using different technique.

On the other hand, GSLT of a system containing interacting DE with
dark matter (DM) enclosed by either apparent or event horizons has
been discussed widely in view of usual entropy of horizon. Karami et
al.$^{{19})}$ have investigated the validity of GSLT on the apparent
and event horizons in non-flat FRW universe. Also, it was
found$^{{20})}$ that the GSLT always holds for a system having
interaction of DE with DM and radiation in non-flat universe
enclosed by apparent horizon. In the discussion of GSLT, it is
realized that the logarithmic and power law corrections to the usual
entropy-area relation of the expanding universe may provide better
results. In this context, some authors$^{{21-23})}$ have explored
the validity of GSLT in closed modified Friedmann equations by using
apparent horizon. A detailed review of DE phenomenon and
thermodynamics in modified gravity from various cosmological aspects
has been provided in the recent work $^{{24})}$.

Recently, Jamil et al.$^{{25})}$ have studied the GSLT of phantom
fluid accreting onto the Schwarzschild BH on the apparent and event
horizons by using logarithmic and power law corrected entropies in
flat FRW universe. Also, some people$^{{26,27})}$ have explored GSLT
in flat universe containing $n$-components of fluid comprising
interactions with corrected entropies of horizon of the universe by
using apparent and event horizons. Here, we extend the
work$^{{25-27})}$ in the closed FRW universe and modified
generalized chaplygin gas (MGCG).

The paper is organized as follows: In section \textbf{2}, we discuss
the GSLT for a system containing the Schwarzschild BH, phantom DE
and also the corrected entropy on the apparent and event horizons.
Section \textbf{3} is devoted to find GSLT constraints of
phantom-like MGCG accretion onto the Schwarzschild BH with
logarithmic and power law corrected entropies of horizon. In section
\textbf{4}, we investigate the GSLT for a system interacting
$n$-components of fluid with logarithmic and power law corrected
entropies on the apparent and event horizons. Section \textbf{5}
explores the validity of GSLT for thermal non-equilibrium case. In
the last section, we summarize our results.

\section{Schwarzschild BH in Phantom Cosmology and GSLT}

In this section, we discuss the GSLT of the system enclosed by
apparent as well as event horizons in the closed universe. The
non-flat FRW universe is given by
\begin{equation}\label{1}
ds^{2}=-dt^{2}+a^{2}(t)[\frac{dr^{2}}{1-kr^{2}}
+r^{2}(d\theta^{2}+\sin^2\theta d\phi^{2})].
\end{equation}
Here $a(t)$ is the cosmic scale factor which measures expansion of
the universe and $k=-1,0,1$ represents the spatial curvature
indicating the open, flat and closed universes, respectively, but we
only assume the closed case. The corresponding equations of motion
are
\begin{eqnarray}\label{2}
H^2+\frac{k}{a^{2}}&=&\frac{8\pi}{3}\rho,\\\label{3}
\dot{H}+H^2&=&-\frac{8\pi}{6}(\rho+3p).
\end{eqnarray}
We assume that the phantom DE accretes onto the Schwarzschild BH.
This phantom DE is described by the perfect fluid whose equation of
continuity is written as
\begin{eqnarray}\label{4}
\dot{\rho}_{ph}+3H(\rho_{ph}+p_{ph})=0.
\end{eqnarray}
Here $\rho_{ph}$ and $p_{ph}$ indicate the energy density and
pressure of the phantom energy that do not obey the null energy
condition, i.e., $\rho_{ph}+p_{ph}<0$. Babichev et al.$^{{6})}$
found that the phantom energy causes the reduction of the
Schwarzschild BH mass. They calculated the following rate of change
in the BH mass
\begin{equation}\label{5}
\dot{M}=\pi Zr^2_{bh}(\rho_{ph}+p_{ph}),
\end{equation}
where $Z$ is a positive dimensionless constant and $r_{bh}=2M$
represents radius of the Schwarzschild BH.

Now we would like to check the validity of the GSLT of the system
which contains BH and phantom energy enclosed by apparent and future
event horizons. The GSLT is given in the form
\begin{equation}\label{6}
\dot{S}_{tot}\equiv\dot{S}_{BH}+\dot{S}_{ph}+\dot{S}_{h}\geq0,
\end{equation}
where $S_{BH}~,S_{ph}$ and $S_{h}$ show entropies due to BH horizon,
phantom fluid and cosmological horizon, respectively. In order to
discuss the GSLT,  we need Gibb's relation
\begin{equation}\label{7}
TdS=d(pV)+pdV,
\end{equation}
where $T,~S$ denote temperature and entropy of the phantom fluid
respectively, and $V=\frac{4}{3}\pi R^3_h$ ($R_h$ as the radius of
the cosmological horizon) is the volume of the spherical system .
Differentiating Eq.(\ref{7}) with respect to time and using
expression of volume, we get
\begin{equation}\label{8}
\dot{S}_{ph}=2\pi
H^2\left(\frac{\dot{H}}{H^2}-\Omega_k\right)R^3_h(HR_h-\dot{R}_h),
\end{equation}
where we have used temperature of the fluid as $T=\frac{1}{2\pi
R_h}$ and $\Omega_k=\frac{k}{a^2H^2}$.

There exist two types of horizon in the accelerated expansion of the
universe for which we can associate entropy for collecting
information behind them. The apparent horizon is the most natural
choice given by$^{{28})}$
\begin{equation}\label{9}
R_a=\frac{1}{\sqrt{H^2+\frac{k}{a^2}}}.
\end{equation}
The event horizon is another choice of cosmological horizon defined
as$^{{29})}$
\begin{equation}\label{10}
R_e=a(t)\int^{\infty}_{a}{\frac{d\tilde{a}}{H\tilde{a}^{2}}}
=a(t)\int^{\infty}_{t}{\frac{d\tilde{t}}{a(\tilde{t})}}.
\end{equation}
It has more conceptual resemblance with the BH horizon. This horizon
has attained much attention after description of the holographic DE
with this horizon. For the sake of generality, we use these two
horizons in our study. Differentiation of the above
equation leads to
\begin{equation}\label{11}
\dot{R_e}=HR_e-1.
\end{equation}

\subsection{Quantum Corrections to Entropy}

The entropy-area relation proposed by Bekenstein$^{{30})}$ is
\begin{equation}\label{10r}
S_{bh}=\frac{k_bA}{4l^2_{pl}},
\end{equation}
where $k_b,~A$ and $l_{pl}=\sqrt{G\hbar c^3}$ represent Boltzmann
constant, horizon area of BH and Planck's length respectively. Here
we take $c=G=\hbar=l_{pl}=k_b=1$. We use $A=4\pi r^2_{bh}$ as the
area of BH horizon and $A=4\pi R^2_h$ (where $R_h$ is either
apparent horizon or event horizon) as the area of cosmological
horizon (outer most boundary of cosmological system) and $R_h\gg
r_{bh}$.

The logarithmic correction to Bekenstein's relation is also obtained
in the scenario of loop quantum gravity. This correction is found
through infinite series expansion of entropy-area relation as
\begin{eqnarray*}
S_{h}&=&\frac{A}{4\hbar}+\tilde{\beta}\ln\left(\frac{A}{4\hbar}\right)-
\tilde{\beta_1}\left(\frac{4\hbar}{A}\right)-\tilde{\beta_2}\left(\frac{16\hbar^2}{A^2}\right)-...\\\nonumber
&=&S_{0}+\tilde{\beta}\ln
S_{0}-\sum^{\infty}_{l=1}\frac{\tilde{\beta_l}}{S^{l}_{0}},
\end{eqnarray*}
where $\tilde{\beta_l}$ appear as finite constants, $S_{0}$
represents the usual entropy of BH, while the remaining terms
involve as a quantum corrections. For cosmological analysis, the
first order correction is worthy and its higher order corrections
can be ignored due to smallness of $\hbar$. Thus, we assume first
order logarithmic corrected entropy$^{{31-33})}$
\begin{equation}\label{12}
S_{h}=\frac{A}{4}+\mu\ln\left(\frac{A}{4}\right)+\nu,
\end{equation}
where $\mu$ and $\nu$ are constants. Cai et al.$^{{34})}$ found that
the logarithmic term involves in a model of entropic cosmology
compensates the framework of the early inflation as well as present
acceleration.

In the scenario of entanglement of entropy in the quantum fields, it
is required to modify the entropy-area relation$^{{35})}$. The
entanglement entropy is stated as \textit{a measure of the
information loss due to the spatial separation between the degrees
of freedom inside and outside the horizon}. It is observed that
there exist several modes of gravitational fluctuations in the
surroundings of BH which behaves like scalar fields and its
entanglement entropy can be found by tracing degrees of freedom
inside the horizon. Also, the authors$^{{35})}$ showed that the
Bekenstein's relation of entropy works only in the ground state. In
case of superposition of ground and excited states of a field, a
correction in term of power law form is included as$^{{35})}$
\begin{equation}\label{13}
S_{h}=\frac{A}{4}\left[1-B_{\lambda}A^{1-\frac{\lambda}{2}}\right],
\end{equation}
where
$B_{\lambda}=\frac{\lambda(4\pi)^{\frac{\lambda}{2}-1}}{(4-\lambda)r^{2-\lambda}_c}$,
$\lambda$ is a constant and $r_c$ is the crossover scale. Firstly,
we check the validity of GSLT for these two corrections on apparent
and future event horizons.

\subsubsection*{Logarithmic Entropy Correction}

The entropy of BH in logarithmic form becomes
\begin{equation}\label{14}
S_{BH}=4\pi M^2+\mu\ln(4\pi M^2)+\nu,
\end{equation}
here we use $r_h=2M$ in the horizon area of black hole. The rate of
change of BH entropy with logarithmic correction is obtained through
Eqs.(\ref{2}), (\ref{4}), (\ref{5}) and (\ref{14})
\begin{eqnarray}\label{15}
\dot{S}_{BH}=-8ZMH^2\left(\frac{\dot{H}}{H^2}-\Omega_k\right) [4\pi
M^2+\mu],
\end{eqnarray}
while for the horizon entropy of the universe on the apparent
horizon, it follows that
\begin{eqnarray}\label{16}
\dot{S}_{A}=\frac{2\pi}{H(1+\Omega_{k})^2}\left(-\frac{\dot{H}}{H^2}+\Omega_k\right)
\left[1+\frac{\mu H^2(1+\Omega_{k})}{\pi}\right].
\end{eqnarray}
It is noted that there are some possibilities for which
$\dot{S}_{A}\leq0$, i.e., $\mu\leq0$ with
$\mu\leq-\frac{\pi}{H^2(1+\Omega_{k})}$ in quintessence case
($\dot{H}<0$). However, two sets of conditions exist for phantom
case ($\dot{H}>0$) which are
$\Omega_{k}\geq\frac{\dot{H}}{H^2},~\mu<0,~\mu\geq-\frac{\pi}{H^2(1+\Omega_{k})}$
and $\Omega_{k}\leq\frac{\dot{H}}{H^2},~\mu\geq0$. Using
Eqs.(\ref{8}), (\ref{9}), (\ref{15}) and (\ref{16}), we have the
total change in entropy on the apparent horizon as
\begin{eqnarray}\nonumber
\dot{S}_{tot}&=&\frac{2\pi}{H(1+\Omega_{k})^3}\left(-\frac{\dot{H}}{H^2}+\Omega_k\right)
\left[\frac{\mu
H^2(1+\Omega_{k})^2}{\pi}+\frac{4ZMH^3}{\pi}\right.\\\label{17}
&\times&\left.(1+\Omega_{k})^3(4\pi
M^2+\mu)+\left(-\frac{\dot{H}}{H^2}+\Omega_k\right)\right].
\end{eqnarray}

It has been found by Sadjadi$^{{16})}$ that the derivative of the
Hubble parameter is positive for phantom dominated universe and
negative for quintessence dominated universe. Since we have
considered phantom dominated universe, so $\dot{H}>0$. Thus the GSLT
is respected for $\Omega_k>\frac{\dot{H}}{H^2}$ with $\mu\geq0$ or
$\mu\leq0$ and
\begin{eqnarray}\nonumber
\frac{\dot{H}}{H^2}\leq\frac{\mu
H^2(1+\Omega_{k})^2}{\pi}+\frac{4ZM_cH^3}{\pi}(1+\Omega_{k})^3(4\pi
M_c^2+\mu)+\Omega_k,
\end{eqnarray}
where $M_c$ is the critical mass of BH. Also,
the GSLT follows for $\Omega_k<\frac{\dot{H}}{H^2}$ with
$\mu\leq0,~M_c<\sqrt{-\frac{\mu}{4\pi}}$ or $\mu\geq0$ with
\begin{eqnarray}\nonumber
\frac{\dot{H}}{H^2}\geq\frac{\mu
H^2(1+\Omega_{k})^2}{\pi}+\frac{4ZM_cH^3}{\pi}(1+\Omega_{k})^3(4\pi
M_c^2+\mu)+\Omega_k.
\end{eqnarray}

The rate of change of entropy of the phantom fluid and horizon
entropy of the universe on the event horizon are
\begin{eqnarray}\label{18}
\dot{S}_{ph}&=&2\pi H^2R^3_e\left(\frac{\dot{H}}{H^2}-\Omega_k\right),\\\label{19}
\dot{S}_{e}&=&2\pi(HR_e-1)\left(R_e+\frac{\mu}{\pi
R_e}\right).
\end{eqnarray}
It is observed that $\dot{S}_{e}\leq0$ for phantom dominated case
with $\mu\geq0$ while for quintessence case with $\mu\leq-\pi
R^2_e$.

Combining Eqs.(\ref{15}), (\ref{18}) and (\ref{19}), the total
entropy turns out to be
\begin{eqnarray}\nonumber
\dot{S}_{tot}&=&2\pi(HR_e-1)\left(R_e+\frac{\mu}{\pi
R_e}\right)-2\left(\frac{\dot{H}}{H^2}-\Omega_k\right)[4ZMH^2(4\pi M^2\\\label{20}
&+&\mu)-\pi H^2R^3_e].
\end{eqnarray}
In the quintessence region, $\dot{H}<0$ and $\dot{R}_e>0$
($HR_e>1$), but this case is skipped because we consider phantom
fluid as mentioned earlier. In the phantom region, $\dot{H}>0$ and
$\dot{R}_e<0$, i.e., $HR_e<1$, it is identified that the GSLT is
satisfied under two possible choices for a closed universe as
\begin{enumerate}
\item $\mu\geq0$ with
\begin{eqnarray}\nonumber
\dot{R}_e\geq\frac{\left(\frac{\dot{H}}{H^2}-\Omega_k\right)R_e}{\pi R^2_e+\mu}
[2ZM_cH^2(4\pi M_c^2+\mu)-\pi H^2R^3_e].
\end{eqnarray}
\item $-\pi R^2_e\leq\mu\leq0$ with either $M_c<\sqrt{-\frac{\mu}{4\pi}},~\Omega_k<\frac{\dot{H}}{H^2}$
or $\Omega_k>\frac{\dot{H}}{H^2},\\\mu\geq\frac{\pi
R^3_e}{2ZM_c}-4\pi M^2_c$.
\end{enumerate}

\subsubsection*{Power Law Entropy Correction}

In the similar way, we can evaluate the rates of change of BH
entropy and horizon entropy of the universe (on the apparent
horizon) with power law correction as
\begin{eqnarray}\label{21}
\dot{S}_{BH}&=&-32\pi
ZM^3H^2\left(\frac{\dot{H}}{H^2}-\Omega_k\right)
\left[1-\frac{\lambda}{2}\left(\frac{2M}{r_{c}}\right)^{2-\lambda}\right],\\\label{22}
\dot{S}_{A}&=&\frac{2\pi}{H(1+\Omega_k)^2}
\left(\frac{\dot{H}}{H^2}-\Omega_k\right)\left[1-\frac{\lambda}{2}
(r_{c}H\sqrt{1+\Omega_k})^{\lambda-2}\right].
\end{eqnarray}
This rate of change of entropy turns out to be negative for
$\lambda\leq0$ in quintessence case ($\dot{H}<0$). However, two sets
of condition exist for phantom case ($\dot{H}>0$), i.e.,
$\Omega_{k}\geq\frac{\dot{H}}{H^2},~\lambda\leq0$ and
$\Omega_{k}\leq\frac{\dot{H}}{H^2},~\mu\geq0,~\lambda\geq\frac{1}{r_{c}H\sqrt{1+\Omega_k}}$.
In view of Eqs.(\ref{8}), (\ref{9}), (\ref{21}) and (\ref{22}), the
rate of change of total entropy on the apparent horizon leads to
\begin{eqnarray}\nonumber
\dot{S}_{tot}&=&\frac{\pi}{H(1+\Omega_k)^2}
\left(\frac{\dot{H}}{H^2}-\Omega_k\right)\left[\lambda
(r_{c}H\sqrt{1+\Omega_k})^{\lambda-2}-8M^3ZH^3\right.\\\label{23}
&\times&\left(1-\frac{\lambda}{2}\left(\frac{2M}{r_{c}}\right)^{2-\lambda}\right)(1+\Omega_k)^2
+\left.\frac{2}{1+\Omega_k}\left(\frac{\dot{H}}{H^2}-\Omega_k\right)\right].
\end{eqnarray}
This expression remains positive, i.e., $\dot{S}_{tot}\geq0$ in the phantom regime as
$\Omega_k>\frac{\dot{H}}{H^2}$ with either $\lambda\leq0$ or
$\lambda\geq0$ and
\begin{eqnarray}\nonumber
\frac{\dot{H}}{H^2}<-\frac{\lambda}{2}
(r_{c}H)^{\lambda-2}(1+\Omega_k)^\lambda+4M_c^3ZH^3(1+\Omega_k)\left(1-\frac{\lambda}{2}
\left(\frac{2M_c}{r_{c}}\right)^{2-\lambda}\right)+\Omega_k.
\end{eqnarray}
Also, for $\Omega_k<\frac{\dot{H}}{H^2}$ with either $\lambda\geq0,~M_c\geq\frac{r_c}{2}
\left(\frac{2}{\lambda}\right)^{\frac{1}{\lambda-2}}$
or $\lambda\leq0$ and
\begin{eqnarray}\nonumber
\frac{\dot{H}}{H^2}>-\frac{\lambda}{2}
(r_{c}H)^{\lambda-2}(1+\Omega_k)^\lambda+4M_c^3ZH^3(1+\Omega_k)
\left(1-\frac{\lambda}{2}\left(\frac{2M_c}{r_{c}}\right)^{2-\lambda}\right)
+\Omega_k.
\end{eqnarray}

For the event horizon, the rate of change of horizon entropy of the
universe becomes
\begin{equation}\label{24}
\dot{S}_e=2\pi
R_e\dot{R}_e\left[1-\frac{\lambda}{2}\left(\frac{R_{e}}{r_{c}}\right)^{2-\lambda}\right].
\end{equation}
This rate of change of horizon entropy would be negative for phantom
case with $\lambda\leq0$.  With the help of Eqs.(\ref{18}),
(\ref{21}) and (\ref{24}), we obtain
\begin{eqnarray}\nonumber
\dot{S}_{tot}&=&2\pi
R_e(HR_e-1)\left[1-\frac{\lambda}{2}\left(\frac{R_{e}}{r_{c}}\right)^{2-\lambda}\right]
-2\left(\frac{\dot{H}}{H^2}-\Omega_k\right)\\\label{25}
&\times&\left[4\pi
ZM^3H^2\left(1-\frac{\lambda}{2}\left(\frac{2M}{r_{c}}\right)^{2-\lambda}\right)-\pi
R^3_eH^2\right].
\end{eqnarray}
It turns out that the GSLT is respected in the phantom region when
\begin{enumerate}
\item $\lambda\geq0$ with either $\Omega_k>\frac{\dot{H}}{H^2},~r_c\geq(\frac{2}{\lambda})^{-2+\lambda}R_e,
~r_c\geq(\frac{2}{\lambda})^{-2+\lambda}M_c$ or
$\Omega_k<\frac{\dot{H}}{H^2},~2(\frac{R_{e}}{r_{c}})^{2-\lambda}\leq\lambda\leq2(\frac{2M_c}{r_{c}})^{2-\lambda}$.
\item $\lambda\leq0$ with
\begin{eqnarray}\nonumber
\dot{R}_e\geq\frac{2\left(\frac{\dot{H}}{H^2}-\Omega_k\right)\left[4\pi
ZM_c^3H^2\left(1-\frac{\lambda}{2}(\frac{2M_c}{r_{c}})^{2-\lambda}\right)-\pi
R^3_eH^2\right]}{2\pi
R_e\left[1-\frac{\lambda}{2}(\frac{R_{e}}{r_{c}})^{2-\lambda}\right]}.
\end{eqnarray}
\end{enumerate}

\section{Accretion of Phantom-like Modified Generalized Chaplygin Gas and GSLT}

In this section, we extend the work of $^{{25})}$ with the most
general EoS of MGCG$^{{36,37})}$ which is stated as
\begin{eqnarray}\label{26}
p=b_1\rho-\frac{b_2}{\rho^m},
\end{eqnarray}
where $b_1,~b_2$ and $m$ are constant parameters. For $b_1=0$, it
reduces to the generalized chaplygin gas, while we can recover the
usual chaplygin gas by setting $b_1=0$ and $m=1$. We assume that a
phantom fluid is described by MGCG and a BH containing in a comoving
volume $V\propto a^3$. The MGCG violates the null energy condition
for $\rho^{1+m}<\frac{b_2}{1+b_1}$. However, the presence of BH
creates distortions in the spacetime inside the cavity, which can be
neglected up to first approximation. Here, we take the total entropy
of the system in the form of logarithmic and power law corrections
to BH entropy with the entropy of the phantom fluid.

\subsubsection*{Logarithmic Entropy Correction}

Using the EoS of MGCG in continuity equation (\ref{4}), we get new
form of entropy as a sum of BH entropy and entropy of phantom energy
\begin{eqnarray}\label{27}
S_n=[4\pi M^2+\mu \log(4\pi
M^2)+\nu]+\left[\frac{1}{u}\left(\rho^{1+m}-\frac{b_2}{1+b_1}\right)^\frac{1}{(1+b_1)(1+m)}V\right],
\end{eqnarray}
where $u$ is a constant of integration. Moreover, in a small
interval of time $\Delta t$, the BH mass changes by an amount
$\Delta M$ due to accretion process as well as change in energy
density of the phantom fluid takes place. These variations lead to
\begin{eqnarray}\nonumber
\Delta S_n&=&8\pi M\left(1+\frac{\mu}{4\pi M^2}\right)\Delta
M+\frac{\rho^{m}}{u(1+b_1)}\left[-\frac{b_2}{u(1+b_1)}\right.\\\label{28}
&+&\left.u^{-1}\rho^{1+m}\right]^\frac{1-(1+b_1)(1+m)}{(1+b_1)(1+m)}V\Delta\rho.
\end{eqnarray}
Notice that the homogeneous and time dependent scalar field with
Lagrangian containing negative kinetic energy term can also
interpret phantom energy. In terms of this scalar field model, the
energy conservation is$^{{18})}$
\begin{eqnarray}\label{29}
\Delta M=-\frac{1}{2}\Delta
\dot{\phi}^2V=-\frac{1}{2\rho^{1+m}}\left((1+b_1)\rho^{1+m}+(1+m)b_2\right)V\Delta
\rho.
\end{eqnarray}
Inserting $V\Delta \rho$ from this relation in Eq.(\ref{28}), we get
\begin{eqnarray}\nonumber
\Delta S_n&=&\left[8\pi M\left(1+\frac{\mu}{4\pi
M^2}\right)-\frac{2\rho^{2m+1}}{u(1+b_1)((1+b_1)\rho^{1+m}+(1+m)b_2)}\right.\\\label{30}
&\times&\left.\left(u^{-1}\rho^{1+m}-\frac{b_2}{u(1+b_1)}\right)^\frac{1-(1+b_1)(1+m)}{(1+b_1)(1+m)}\right]\Delta
M.
\end{eqnarray}

On account of accretion of phantom energy, the mass of BH decreases
as $\Delta M<0$. For the validity of the GSLT, $\Delta S_n>0$, we
put
\begin{eqnarray}\nonumber
4\pi M_c\left(1+\frac{\mu}{4\pi M_c^2}\right)-\delta(\rho)\leq0,
\end{eqnarray}
where
\begin{eqnarray}\nonumber
\delta(\rho)=\frac{2u^{-1}(1+b_1)^{-1}\rho^{2m+1}}{((1+b_1)\rho^{1+m}
+(1+m)b_2)}\left(u^{-1}\rho^{1+m}-\frac{b_2}{u(1+b_1)}\right)^\frac{1-(1+b_1)(1+m)}{(1+b_1)(1+m)}.
\end{eqnarray}
This leads to the following range of the critical mass of BH
\begin{eqnarray}\nonumber
M_{-}\leq M_c\leq M_{+},
\end{eqnarray}
where
\begin{eqnarray}\nonumber
M_{\pm}=\frac{1}{8\pi}\left[\delta(\rho)\pm\sqrt{\delta^2(\rho)-16\pi\mu}\right],
\end{eqnarray}
indicating that the critical mass is bounded above and below by
$M_{+}$ and $M_{-}$ respectively. This also gives information about
accretion process onto a BH that it cannot accrete above and below
the critical values. It is observed that the range of the critical
mass is physically acceptable for $\mu\leq0$ or $\mu\geq0$ with
$\delta^2(\rho)>16\pi\mu$. If $\mu<<\frac{1}{16\pi}\delta^2(\rho)$,
then the critical mass condition turns out to be
$M_c\leq\frac{\delta(\rho)}{4\pi}$.

\subsubsection*{Power Law Entropy Correction}

In this case, the variation in the total entropy becomes
\begin{eqnarray}\label{31}
\Delta S_n=\left[4\pi
M\left(1-\frac{\lambda(2M)^{2-\lambda}}{2r^{2-\lambda}_c}\right)
-\delta(\rho)\right]\Delta M.
\end{eqnarray}
The validity of GSLT leads to the constraint
\begin{eqnarray}\label{32}
M_c\left(1-\frac{\mu(2M_c)^{2-\lambda}}{2r^{2-\lambda}_c}\right)-\frac{\delta(\rho)}{4\pi}<0.
\end{eqnarray}
It is a complicated equation in $M_c$, so we handle it iteratively
for different values of $\lambda=1,~2,~3,~4,~5$.
\begin{itemize}
\item By inserting $\lambda=1$ in Eq.(\ref{32}), we obtain
\begin{eqnarray}\nonumber
M^2_c-r_cM_c+\frac{r_c\delta(\rho)}{4\pi}\geq0.
\end{eqnarray}
It gives the range of the critical mass of BH, i.e.,
$M_{-}\leq M_c\leq M_{+}$, where
\begin{eqnarray}\nonumber
M_{\pm}=\frac{r_c}{2}\left[1\pm\sqrt{1-\frac{r_c\delta(\rho)}{\pi}}\right].
\end{eqnarray}
Here, the critical values are physically acceptable if
$r_c<\frac{\pi}{\delta(\rho)}$.
\item For $\lambda=2$, we have $\frac{\delta(\rho)}{4\pi}\geq0$, which is ruled out as we
cannot find any critical value of the BH mass.
\item For $\lambda=3$, we obtain upper bound of the critical mass of BH
\begin{eqnarray}\nonumber
M_c<\frac{3r_c}{4}+\frac{\delta(\rho)}{4\pi}.
\end{eqnarray}
\item For $\lambda=4$, it follows that
\begin{eqnarray}\nonumber
M^2_c-\frac{\delta(\rho)}{4\pi}M_c-\frac{r^2_c}{2}<0.
\end{eqnarray}
It gives the following range of the critical mass of BH
\begin{eqnarray}\nonumber
M_{-}\leq M_c\leq M_{+}
\end{eqnarray}
where $M_{\pm}=\frac{1}{8\pi}\left[\delta(\rho)
\pm\sqrt{\delta^2(\rho)-32\pi^2r^2_c}\right]$. The roots are real if
$r^2_c<\frac{\delta(\rho)}{32\pi^2}$.
\item For $\lambda=5$, Eq.(\ref{32}) turns out to be
$M^3_c-pM^2_c-q<0$ with
\begin{eqnarray}\nonumber
p=\frac{\delta(\rho)}{4\pi},\quad q=\frac{5r^3_c}{16}.
\end{eqnarray}
To solve this cubic equation, we introduce the variable
$x=M_c-\frac{p}{3}$, which leads to
\begin{eqnarray}\nonumber
x^3+lx+n<0,\quad l=-\frac{p^2}{3},\quad n=-(q+\frac{2}{27}p^3).
\end{eqnarray}
This cubic equation has the following three distinct roots because
$l<0$
\begin{eqnarray}\nonumber
x_1&=&2\sqrt{-\frac{l}{3}}\cos(\varphi), \quad
x_2=2\sqrt{-\frac{l}{3}}\cos(\frac{2\pi}{3}-\frac{\varphi}{3}),\\\nonumber
x_3&=&2\sqrt{-\frac{l}{3}}\cos(\frac{2\pi}{3}+\frac{\varphi}{3}),
\end{eqnarray}
where
\begin{eqnarray}\nonumber
\varphi=\frac{1}{3}\arccos\left(\frac{27q+2p^3}{2p^3}\right).
\end{eqnarray}
Hence there exist two possibilities for the range of critical mass
\begin{eqnarray}\nonumber
M_c\leq M_{1},\quad M_2<M_c<M_3,
\end{eqnarray}
where
\begin{eqnarray}\nonumber
M_1&=&\frac{\delta(\rho)}{6\pi}\left(\frac{1}{2}+\cos(\varphi)\right),\quad
M_2=\frac{\delta(\rho)}{6\pi}\left(\frac{1}{2}+\cos(\frac{2\pi}{3}+\frac{\varphi}{3})\right),\\\nonumber
M_3&=&\frac{\delta(\rho)}{6\pi}\left(\frac{1}{2}+\cos(\frac{2\pi}{3}-\frac{\varphi}{3})\right).
\end{eqnarray}
\end{itemize}

\section{Interacting Scenario and GSLT with Thermal Equilibrium}

Here we consider perfect fluid of $n$-components such as DE, DM,
radiation and so on that exist in the closed universe. Also, $\rho$
and $p$ represent the total energy density and pressure of the
combined fluid, i.e., $\rho_t=\sum^{n}_{i=1}\rho_i$ and
$p_t=\sum^{n}_{i=1}p_i$. We can rewrite Eq.(\ref{2}) in the form of
fractional energy densities as
\begin{equation}\nonumber
\Omega_{t}=1+\Omega_{k},
\end{equation}
where $\Omega_{t}=\frac{\rho_{t}}{3H^{2}}$ and the equation of
continuity in this scenario becomes
\begin{equation}\label{33}
\dot{\rho_t}+3H(\rho_t+p_t)=0.
\end{equation}
We assume the interacting scenario of all fluid components$^{{38})}$
but we are not introducing the specific form of interaction due to
unknown nature of DE and DM. The interaction between the fluid
components changes the nature of the corresponding equations of
state. Moreover, the interaction makes possible transition from a
quintessence state to phantom era and also helps in explaining the
cosmic coincidence problem$^{{39-41})}$. It has been
obtained$^{{42,43})}$ that the interacting DE models favor $95$\%
confidence limits with the data of the observational constraints.
Thus Eq.(\ref{4}) takes the following form for interacting fluid
components
\begin{equation}\label{34}
\dot{\rho}_{i}+3H(\rho_{i}+p_{i})=\Upsilon_i,\quad i=1,2,3,...,n,
\end{equation}
where $\Upsilon_i$ indicate the interacting parameters which may be
taken in terms of Hubble parameter or energy density of DM or DE or
in terms of both$^{{44,45})}$. This term plays the role of energy
exchange between the components of perfect fluid and also it obeys
$\sum^{n}_{i=1}\Upsilon_i=0$.

In this scenario, Gibb's relation takes the form
\begin{equation}\label{35}
T_idS_i=d(p_iV)+p_idV,
\end{equation}
for each component of fluid. Also, $T_i$ and $S_i$ denote
temperature and entropy of the $i$th component of fluid
respectively. Differentiating the above equation with respect to
time and using the expression of volume, we obtain
\begin{equation}\label{36}
\dot{S}_i=\frac{4}{3}\pi R^3_{h}\frac{\Upsilon_{i}}{T_i}+4\pi
R^2_h(\dot{R}_{h}-HR_{h})\frac{\rho_i+p_i}{T_i}.
\end{equation}
Further, the overall variation of entropy inside the horizon takes
the form
\begin{equation}\label{37}
\dot{S}_{I}=\sum^{n}_{i=1}\dot{S}_i=\frac{4}{3}\pi
R^3_{h}\sum^{n}_{i=1}\frac{\Upsilon_{i}}{T_i}+4\pi
R^2_h(\dot{R}_{h}-HR_{h})\sum^{n}_{i=1}\frac{\rho_i+p_i}{T_i}.
\end{equation}
Here, we assume that the temperature of $n$-components of fluid is
the same as the temperature of the horizon, i.e., $\forall~i$,
$T_i=T$ (it is called thermal equilibrium condition). In general, it
is not true because, in the present scenario of the universe,
radiation temperature is larger than the non-relativistic cold DM.

\subsection{Logarithmic Correction to Entropy}

Here we discuss GSLT of the system by taking contribution of
logarithmic correction to entropy into account on the above
mentioned two cosmological horizons.

\subsubsection*{On the Apparent Horizon}

In view of thermal equilibrium, $\sum^{n}_{i=1}\Upsilon_i=0$ and
Hawking temperature, one can get the change in entropy due to
overall fluid inside the apparent horizon as
\begin{equation}\label{38}
\dot{S}_{I}=\frac{2\pi}{H(1+\Omega_{k})^3}\left(\frac{\dot{H}}{H^2}-\Omega_k\right)^2
+\frac{2\pi}{H(1+\Omega_{k})^2}\left(\frac{\dot{H}}{H^2}-\Omega_k\right).
\end{equation}
Consequently, by using Eqs.(\ref{16}) and (\ref{38}), the rate of
change of the total entropy becomes
\begin{equation}\label{39}
\dot{S}_{tot}=\frac{2\pi}{H(1+\Omega_{k})^2}\left(\frac{\dot{H}}{H^2}-\Omega_k\right)
\left[-\frac{\mu
H^2(1+\Omega_{k})}{\pi}+\left(\frac{\dot{H}}{H^2}-\Omega_k\right)\right].
\end{equation}
It is observed that the GSLT always holds for $\mu=0$. We also
discuss two interesting possibilities $\dot{H}<0$ (in quintessence
region) and $\dot{H}>0$ (in phantom region) for the validity of the
GSLT. The condition $\dot{H}<0$ implies that
$\frac{\dot{H}}{H^2}<\Omega_k$ and in this case, the GSLT favors for
either $\mu\geq0$ or $\mu\leq0$ with
$\frac{\dot{H}}{H^2}\leq\frac{\mu H^2(1+\Omega_{k})}{\pi}+\Omega_k$.
The second condition provides two possible choices, i.e.,
$\frac{\dot{H}}{H^2}< \Omega_k$ and $\frac{\dot{H}}{H^2}>\Omega_k$.
If $\frac{\dot{H}}{H^2}<\Omega_k$, then the GSLT follows for either
$\mu\geq0$ or $\mu\leq0$ with $\frac{\dot{H}}{H^2}\leq\frac{\mu
H^2(1+\Omega_{k})}{\pi}+\Omega_k$. However,
$\frac{\dot{H}}{H^2}>\Omega_k$ gives either $\mu\leq0$ or $\mu\geq0$
with $\frac{\dot{H}}{H^2}\geq\frac{\mu
H^2(1+\Omega_{k})}{\pi}+\Omega_k$.

\subsubsection*{On the Event Horizon}

The entropy of the fluid enclosed by the event horizon is
\begin{equation}\label{40}
\dot{S}_{I}=2\pi H^2R^3_e\left(\frac{\dot{H}}{H^2}-\Omega_k\right).
\end{equation}
The total rate of change of entropy can be obtained by
combining Eqs.(\ref{19}) and (\ref{40})
\begin{equation}\label{41}
\dot{S}_{tot}=2\pi(HR_e-1)\left(R_e+\frac{\mu}{\pi R_e}\right)+2\pi
H^2R^3_e\left(\frac{\dot{H}}{H^2}-\Omega_k\right).
\end{equation}
It is observed that the GSLT for logarithmic correction to entropy
with future event horizon in quintessence regime (i.e., $\dot{H}<0$
provided $HR_e>1$) is respected for $\mu\geq0$ with $\dot{R}_e\geq-
\frac{H^2R^3_e\left(\frac{\dot{H}}{H^2}-\Omega_k\right)}{\left(R_e
+\frac{\mu}{\pi R_e}\right)}$ and for $\mu\leq0$ with $\pi
R^2_e\geq-\mu,~\dot{R}_e\geq-
\frac{H^2R^3_e\left(\frac{\dot{H}}{H^2}-\Omega_k\right)}{\left(R_e
+\frac{\mu}{\pi R_e}\right)}$. In phantom regime (i.e., $\dot{H}>0$
provided $HR_e<1$), it yields the following conditions
\begin{enumerate}
\item For $\mu\geq0$ and $\frac{\dot{H}}{H^2}\geq\Omega_k$ with $\dot{R}_e\geq-
\frac{H^2R^3_e\left(\frac{\dot{H}}{H^2}-\Omega_k\right)}{\left(R_e
+\frac{\mu}{\pi R_e}\right)}$, GSLT is satisfied and violated for $\mu>0$ with $\frac{\dot{H}}{H^2}\leq\Omega_k$.
\item For $\mu\leq0$ with $\pi R^2_e\geq-\mu,~\frac{\dot{H}}{H^2}\geq\Omega_k$, it holds.
\end{enumerate}

Moreover, in order to get clear picture of result (\ref{41}), we
consider a model which corresponds to a super accelerated universe
with big rip at a finite time. At this stage, $H$ gets very large
value and this regime is named as phantom dominated universe. This
model is also named as pole-like type scale factor which is given by
\begin{equation}\label{42}
a(t)=a_0(t_s-t)^{-r},\quad a_0>0,\quad r>0,\quad t_s>t.
\end{equation}
By inserting this scale factor in relation (\ref{41}), we have
\begin{equation}\label{43}
\dot{S}_{tot}=\frac{2\pi(t_s-t)}{(r+1)^2}\left[-1-\frac{\mu(r+1)^2}{\pi(t_s-t)^2}+\frac{r}{r+1}(1-r\Omega_k)\right].
\end{equation}
The present time $t=t_0$ with $\Omega_{k_0}=0.01$ provides the
following condition for the validity of GSLT
\begin{equation}\label{44}
t_s-t_0\geq\sqrt{\frac{\pi}{\mu(r+1)^2}\left[1-\frac{r}{r+1}\left(1-\frac{r}{100}\right)\right]}.
\end{equation}
The above result holds either for $\mu>0$ or $\mu<0$ with
$r\geq100$.

\subsection{Power Law Correction to Entropy}

By applying the previous procedure, we discuss the GSLT by
using the horizon entropy with power law correction on the apparent
and event horizons.

\subsubsection*{On the Apparent Horizon}

The total rate of change of entropy on the apparent horizon becomes
\begin{equation}\label{45}
\dot{S}_{tot}=\frac{\pi}{H(1+\Omega_k)^2}
\left(\frac{\dot{H}}{H^2}-\Omega_k\right)\left[\lambda
(r_{c}H\sqrt{1+\Omega_k})^{\lambda-2}+\frac{2}{1+\Omega_k}\left(\frac{\dot{H}}{H^2}-\Omega_k\right)\right].
\end{equation}
Notice that the GSLT always holds for $\lambda=0$. This shows that
if $\dot{H}<0$, then $\frac{\dot{H}}{H^2}<\Omega_k$ and the GSLT
holds for either $\lambda\leq0$ or $\lambda\geq0$ with
\begin{equation*}
\frac{\dot{H}}{H^2}<-\frac{\lambda}{2}(H
r_c)^{\lambda-2}(1+\Omega_k)^{\lambda}+\Omega_k.
\end{equation*}
If $\dot{H}>0$, then there are two possible constraints either
$\frac{\dot{H}}{H^2}<\Omega_k$ or $\frac{\dot{H}}{H^2}>\Omega_k$.
Thus the GSLT follows for $\frac{\dot{H}}{H^2}<\Omega_k$ either
$\lambda\leq0$ or $\lambda\geq0$ with
\begin{equation*}
\frac{\dot{H}}{H^2}<-\frac{\lambda}{2}(H
r_c)^{\lambda-2}(1+\Omega_k)^{\lambda}+\Omega_k.
\end{equation*}
For the second choice, we have constraints either $\lambda\geq0$ or $\lambda\leq0$
with $\frac{\dot{H}}{H^2}>-\frac{\lambda}{2}(H
r_c)^{\lambda-2}(1+\Omega_k)+\Omega_k$. Otherwise, it is violated.

\subsubsection*{On the Event Horizon}

The overall change of entropy takes the form
\begin{equation}\label{46}
\dot{S}_{tot}=2\pi
R_e(HR_e-1)\left[1-\frac{\lambda}{2}\left(\frac{R_{e}}{r_{c}}\right)^{2-\lambda}\right]+
2\pi H^2R^3_e\left(\frac{\dot{H}}{H^2}-\Omega_k\right).
\end{equation}
Here, $\dot{S}_{tot}\geq0$ in the quintessence region as either
$\lambda\geq0$ with
$r_c\geq(\frac{2}{\lambda})^{-2+\lambda}R_e$ and
\begin{equation*}
\dot{R}_e\geq-\frac{2\left(\frac{\dot{H}}{H^2}-\Omega_k\right)\pi
R^3_eH^2}{2\pi
R_e\left[1-\frac{\lambda}{2}(\frac{R_{e}}{r_{c}})^{2-\lambda}\right]}
\end{equation*}
or $\lambda\leq0$ with the above condition. However, in the phantom
region, it has the following validity conditions
\begin{enumerate}
\item $\lambda\geq0$ with either $\Omega_k>\frac{\dot{H}}{H^2},~\lambda\geq2(\frac{R_{e}}{r_{c}})^{2-\lambda}$ or
$\Omega_k<\frac{\dot{H}}{H^2},~2(\frac{R_{e}}{r_{c}})^{2-\lambda}\leq\lambda\leq2(\frac{2M_c}{r_{c}})^{2-\lambda}$.
\item $\lambda\leq0$ with $\dot{R}_e\geq-\frac{2\left(\frac{\dot{H}}{H^2}-\Omega_k\right)\pi
R^3_eH^2}{2\pi
R_e\left[1-\frac{\lambda}{2}(\frac{R_{e}}{r_{c}})^{2-\lambda}\right]}$.
\end{enumerate}
Substituting the scale factor (\ref{42}), we obtain
\begin{equation}\label{47}
\dot{S}_{tot}=\frac{2\pi(t_s-t)}{(r+1)^2}\left[\frac{\lambda(t_s-t)^{2-\lambda}}{2(r_c(r+1))^{2-\lambda}}
-\frac{1}{r+1}(1+r^2\Omega_k)\right].
\end{equation}
The GSLT is valid in the present time as
\begin{equation}\label{48}
t_s-t_0\geq
r_c(r+1)\left[\frac{2}{\lambda(r+1)}\left(1+\frac{r^2}{100}\right)\right]^\frac{1}{2-\lambda},
\end{equation}
which holds for $\lambda>0$.

\section{GSLT with Thermal Non-Equilibrium}

In order to explore the GSLT with non-equilibrium condition, we
restrict the system into two non-interacting components such as DM
and DE for getting more insights. In this scenario, two
non-conserving equations turn out to be
\begin{equation}\label{49}
\dot{\rho}_{DM}+3H(1+\omega_{DM})\rho_{DM}=0,\quad
\dot{\rho}_{DE}+3H(1+\omega_{DE})\rho_{DE}=0.
\end{equation}
Here $\rho_{DM},~\rho_{DE}$ are energy densities and
$\omega_{DM},~\omega_{DE}$ are EoS parameters of DM and DE
respectively. Using Eqs.(\ref{1}) and (\ref{49}), we get the entropy
of fluid in this scenario as
\begin{eqnarray}\nonumber
\dot{S}_{I}&=&R^2_h(\dot{R}_{h}-HR_{h})\left[4\pi(1+\omega_{DM})\rho_{0}a^{-3(1+\omega_{DM})}\left(\frac{1}{T_{DM}}
-2\pi R_h\right)\right.\\\label{50} &-&\left.2\pi
H^2R_h\left(\frac{\dot{H}}{H^2}-\Omega_k\right)\right].
\end{eqnarray}
where $T_{DM}$ represents the temperature of DM while we use
temperature $T_{DE}=2\pi R_{h}$ for DE component of fluid and
$\rho_{0}$ is positive integration constant.

\subsection{Logarithmic Correction to Entropy}

Here we check the validity of GSLT in case of thermal
non-equilibrium on the apparent and event horizons. The rate of
change of total entropy on the apparent horizon can be obtained by
using Eqs.(\ref{16}) and (\ref{50}) as
\begin{eqnarray}\nonumber
\dot{S}_{tot}&=&-\frac{4\pi(1+\omega_{DM})\rho_{0}}{H^2(1+\Omega_k)^{\frac{3}{2}}}
\left(1+\frac{\frac{\dot{H}}{H^2}-\Omega_k}{1+\Omega_k}\right)
\left(\frac{1}{T_{DM}}-\frac{2\pi}{H\sqrt{1+\Omega_k}}\right)a^{-3(1+\omega_{DM})}\\\nonumber
&+&\frac{2\pi}{H(1+\Omega_k)^3}\left(\frac{\dot{H}}{H^2}
-\Omega_k\right)\left[\left(\frac{\dot{H}}{H^2}
-\Omega_k\right)\sqrt{1+\Omega_k}-\frac{\mu}{\pi^2}(1+\Omega_k)^2\right].\\\label{51}
\end{eqnarray}
This shows that GSLT is valid only if
$T_{DM}\geq\frac{H\sqrt{1+\Omega_k}}{2\pi}$ along with
$\frac{\dot{H}}{H^2}\geq\Omega_k$ and $\mu\leq0$. The total rate of
change of entropy on the event horizon turns out to be
\begin{eqnarray}\nonumber
\dot{S}_{tot}&=&-4\pi R^2_e\left(\frac{1}{T_{DM}}-2\pi
R_e\right)(1+\omega_{DM})\rho_{0}a^{-3(1+\omega_{DM})}\\\label{52}
&+&2\pi R_e\left[H^2R^2_e\left(\frac{\dot{H}}{H^2}
-\Omega_k\right)+\dot{R}_e\left(1+\frac{\mu}{\pi R_e}\right)\right].
\end{eqnarray}
Here, GSLT is respected for $T_{DM}\geq\frac{1}{2\pi R_e}$ with
$\frac{\dot{H}}{H^2}\geq\Omega_k,~\dot{R}_e\geq0$ and $\mu\geq0$.

\subsection{Power Law Correction to Entropy}

Now we find the validity conditions for power law correction on the
apparent and event horizons. Equations (\ref{22}) and (\ref{50})
yield the following expression on the apparent horizon
\begin{eqnarray}\nonumber
\dot{S}_{tot}&=&-\frac{4\pi(1+\omega_{DM})\rho_{0}}{H^2(1+\Omega_k)^{\frac{3}{2}}}
\left(1+\frac{\frac{\dot{H}}{H^2}-\Omega_k}{1+\Omega_k}\right)
\left(\frac{1}{T_{DM}}-\frac{2\pi}{H\sqrt{1+\Omega_k}}\right)\\\nonumber
&\times&a^{-3(1+\omega_{DM})}+\frac{2\pi}{H(1+\Omega_k)^3}
\left(\frac{\dot{H}}{H^2}-\Omega_k\right)
\left[\frac{\lambda}{2}(r_{c}H\sqrt{1+\Omega_k})^{\lambda-2}\right.\\\label{53}
&\times&\left.(1+\Omega_k)
+\left(\frac{\dot{H}}{H^2}-\Omega_k\right)\right].
\end{eqnarray}
The GSLT is respected for the same conditions as mentioned after
Eq.(\ref{51}) except $\lambda>0$. In this correction, the total rate
of change of entropy takes the following form
\begin{eqnarray}\nonumber
\dot{S}_{tot}&=&-4\pi R^2_e\left(\frac{1}{T_m}-2\pi
R_e\right)(1+\omega_{DM})\rho_{0}a^{-3(1+\omega_{DM})}\\\label{54}
&+&2\pi R_e\left[H^2R^2_e\left(\frac{\dot{H}}{H^2}
-\Omega_k\right)+\dot{R}_e\left(1-\frac{\lambda}{2}\left(\frac{R_e}{r_c}\right)^{2-\lambda}\right)\right].
\end{eqnarray}
which is positive and increasing for the same conditions as given
after Eq.(\ref{52}) except $\lambda<0$.

\section{Concluding Remarks}

The evidence of spatial curvature in the universe comes from
different observational data that predict the contribution of
spatial curvature to the total energy density of the
universe$^{{46-52})}$. Also, the early inflation era follows the
flat universe but it is only possible for large number of
e-folding$^{{53})}$. The evidences about the non-flat universe was
also occurred through the data of the first year WMAP
analysis$^{{54})}$. Moreover, the parameterizations of DE models
allow the spatial curvature in the universe by using observations of
type Ia supernovae (SNe Ia), baryon acoustic oscillation (BAO) and
CMBR$^{{55})}$. By using WMAP five year data, it is found that
$\Omega_k$ remains in the range $(-0.2851,0.0099)$ at 95\%
confidence level$^{{56})}$. This range is improved to
$(-0.0181,0.0071)$ by incorporating the data of BAO and SNe Ia.
Recently, a little improvement appears in the range of $\Omega_k$ by
using latest WMAP $7$ and $9$-years data$^{{57,58})}$.

The GSLT is a basic ingredient of physics and its validity needs to
be checked for thermal equilibrium and non-equilibrium frameworks in
different cosmological systems (which includes either BH or
n-components of fluid). The inclusion of BH entropy in the
cosmological system is an interesting development. The earlier
works$^{{25-27})}$ on this subject have been done for improving the
shortcomings in the usual horizon entropy so that the GSLT is
checked in general relativity through entropy corrections. These
have been investigated with the help of phantom and quintessence
regions through $\dot{H}$ as suggested by Sadjadi$^{16}$ and set
some physical conditions for the validity of GSLT. For the sake of
consistency of the results, they have used apparent and event
horizons. However, all the works have been done in flat FRW universe
in this area.

The above discussion motivates us to study the GSLT in the closed
FRW universe and we have discussed GSLT in three folds by assuming
quantum corrected entropies to the horizon of the universe
(logarithmic and power law versions). These corrections are
specified by constant parameters $\mu$ and $\nu$ in logarithmic case
and $\lambda$ in power law case that play crucial role in the debate
of results. We have investigated GSLT on apparent and event horizons
for making our study more general.
\begin{itemize}
\item Firstly, we have checked the validity of the GSLT for a system containing the
accretion of phantom fluid onto the Schwarzschild BH on the apparent
and event horizons in closed universe. We have found different
possibilities for the validity of the GSLT for the logarithmic and
power law forms of entropy in this scenario (see after
Eqs.(\ref{17}), (\ref{20}) and Eqs.(\ref{23}), (\ref{25})
respectively). Since the entropy-area relation is same for any
causal horizon, i.e., either it is BH or cosmological, hence for the
sake of consistency, we have chosen quantum corrected entropies of
these horizons.
\item Secondly, we have developed the GSLT for a system containing
only BH and phantom fluid described by MGCG (the most general EoS).
We have found the critical values of the BH mass above which the
phantom fluid cannot accrete onto a BH because of the violation of
the GSLT. This is an interesting phenomenon which relates the
entropy of fluid and mass of the BH. In accretion process, the
entropy of the BH decreases but in order to keep the positivity of
the total entropy, the entropy of fluid must increase enough to
maintain it.
\item Finally, we have investigated the GSLT for a system of
$n$-components of fluid and quantum corrected entropies in the
closed universe on the apparent and event horizons. We have made our
analysis in thermal equilibrium and non-equilibrium (for
non-interacting DM and DE components of fluid) and found the
conditions under which the GSLT is valid. We have discussed the
results by focusing on $\dot{H}$ (with $\lessgtr0$) and also
pole-like scale factor in case of event horizon in equilibrium case.
\end{itemize}
The important description of this work is the inclusion of
fractional curvature density which has led our results to more
general as compared to past works. This density along with entropy
corrections plays an important role in securing the validity of
GSLT. It is also mentioned here that we have developed a thermal
system by introducing MGCG which provides more insights about GSLT
as compared to simple EoS. In conclusion, we have put some reliable
conditions on different cosmological parameters in quintessence and
phantom phases of the universe (for two systems including
\textit{combination of BH and phantom fluid} and
\textit{n-components of fluid} in thermal and non-thermal
equilibrium) for the validity of GSLT. For a system having only BH
and MGCG, we have mentioned lower and upper bounds of BH mass for
which GSLT is respected.

\vspace{0.5cm}

{\bf Acknowledgment}

\vspace{0.5cm}

We would like to thank the Higher Education Commission, Islamabad,
Pakistan for its financial support through the \textit{Indigenous
Ph.D. 5000 Fellowship Program Batch-VII}.

\vspace{0.5cm}

1) S. Perlmtter, G. Aldering, G. Goldhaber, R. A. Knop, P. Nugent,
P. G. Castro, S. deustua, S. Fabbro, A. Goobar, D. E. Groom, I. M.
Hook, A. G. Kim, M. Y. Kim, J. C. Lee, N. J. Nunes, R. Pain, C. R.
Pennypacker and R. Quimby: Astrophys. J. \textbf{517} (1999) 565.\\
2) R. R. Caldwell and M. Doran: Phys. Rev. D \textbf{69} (2004)
103517.\\
3) T. Koivisto and D. F. Mota: Phys. Rev. D \textbf{73} (2006)
083502.\\
4) S. F. Daniel: Phys. Rev. D \textbf{77} (2008) 103513.\\
5) S. Nesseris and L. Perivolaropoulos: Phys. Rev. D \textbf{70}
(2004) 123529.\\
6) E. Babichev, V. Dokuchaev and Y. Eroshenko: Phys. Rev. Lett.
\textbf{93} (2004) 021102.\\
7) A. V. Yurov, P. M. Moruno and P. F. G. Diaz: Nuc. Phys. B
\textbf{759} (2006) 320.\\
8) P. Martin-Moruno: Phys. Lett. B \textbf{659} (2008) 40.\\
9) P. Martin-Moruno, A. L. Marrakchi, S. Robles-Perez and P. F. G.
Diaz: Gen. Relativ. Gravit. \textbf{41} (2009) 2797.\\
10) M. Jamil, M. A. Rashid and A. Qadir: Eur. Phys. J. C \textbf{58}
(2008) 325.\\
11) E. Babichev, S. Chernov, V. Dokuchaev and Y. Eroshenko: Phys. Rev. D \textbf{78} (2008) 104027.\\
12) M. Jamil: Eur. Phys. J. C \textbf{62} (2009) 325.\\
13) J. Bhadra and U. Debnath: Eur. Phys. J. C \textbf{72} (2012)
1912.\\
14) G. Izquierdo and D. Pav$\acute{o}$n: Phys. Lett. B \textbf{633}
(2006) 420.\\
15) G. Izquierdo and D. Pav$\acute{o}$n: Phys. Lett. B \textbf{639}
(2006) 1.\\
16) H. M. Sadjadi: Phys. Rev. D \textbf{73} (2006) 063525.\\
17) H. M. Sadjadi: Phys. Lett. B \textbf{645} (2007) 108.\\
18) D. F. J. A. Pacheco and J. E. Horvath: Class. Quantum Grav.
\textbf{24} (2007) 5427.\\
19) K. Karami, S. Ghaffari and M. M. Soltanzadeh: Class. Quantum
Grav. \textbf{27} (2010) 205021.\\
20) M. Jamil, E. N. Saridakis and M. R. Setare: Phys. Rev. D
\textbf{81} (2010) 023007.\\
21) N. Radicella and D. Pav$\acute{o}$n: Phys. Lett. B \textbf{704}
(2011) 260.\\
22) K. Karami, A. Abdolmaleki, N. Sahraei and S. Ghaffari: JHEP
\textbf{150} (2011) 1108.\\
23) K. Karami, A. Sheykhi, N. Sahraei and S. Ghaffari: Eur. Phys.
Lett. \textbf{93} (2011) 29002.\\
24) K. Bamba, S. Capozziello, S. Nojiri, S. D. Odintsov: Astrophys.
Space Sci. \textbf{337} (2012) 789.\\
25) M. Jamil, D. Momeni, K. Bamba and R. Myrzakulov: Int. J. Mod.
Phys. D \textbf{21} (2012) 1250065.\\
26) H. M. Sadjadi and M. Jamil: Eur. Phys. Lett. \textbf{92} (2010)
69001.\\
27) U. Debnath, S. Chattopadhyay, I. Hussain, M. Jamil and R.
Myrzakulov: Eur. Phys. J. C \textbf{72} (2012) 1875.\\
28) D. Bak and S. J. Rey: Class. Quantum Grav. \textbf{17} (2000)
L83.\\
29) M. Li: Phys. Lett. B \textbf{603} (2004) 1.\\
30) J. D. Bekenstein: Phys. Rev. D \textbf{9} (1974) 3292.\\
31) R. Banerjee and S. K. Modak:  JHEP \textbf{073} (2009) 0911.\\
32) H. Wei: Commun. Theor. Phys. \textbf{52} (2009) 743.\\
33) S. Banerjee, R. K. Gupta and A. Sen: JHEP \textbf{147} (2011) 1103.\\
34) Y. F. Cai, J. Liu and H. Li: Phys. Lett. B \textbf{690} (2010)
213.\\
35) S. Das, S. Shankaranarayanan and S. Sur: Phys. Rev. D
\textbf{77} (2008) 064013.\\
36) H. B. Benaoum: arXiv: 0205140.\\
37) Debnath, U., A. Banerjee and S. Chakraborty: Class. Quantum Grav. \textbf{21} (2004) 5609.\\
38) W. Zimdahl: Int. J. Mod. Phys. D \textbf{14} (2005) 2319.\\
39) M. R. Setare: Eur. Phys. J. C \textbf{50} (2007) 991.\\
40) K. Karami, S. Ghaffari and J. Fehri: Eur. Phys. J. C \textbf{64}
(2009) 85.\\
41) H. M. Sadjadi and M. Jamil: Gen. Relativ. Gravit. \textbf{43}
(2011) 1759.\\
42) J. Valiviita, R. Maartens and E. Majerotto: Mon. Not. R. Astron.
Soc. \textbf{402} (2010) 2355.\\
43) H. Wei: Phys. Lett. B \textbf{691} (2010) 173.\\
44) H. M. Sadjadi and M. Honardoost: Phys. Lett. B \textbf{647}
(2007) 231.\\
45) H. M. Sadjadi and M. Alimohammadi: Phys. Rev. D
\textbf{74} (2006) 103007.\\
46) J. L. Sievers, J. R. Bond, J. K. Cartwright, C. R. Contaldi, B.
S. Mason, S. T. Myers, S. Padin, T. J. Pearson, U.L. Pen, D.
Pogosyan, S. Prunet, A. C. S. Readhead, M. C. Shepherd, P. S.
Udomprasert, L. Bronfman, W. L. Holzapfel, and J. May: Astrophys. J.
\textbf{591} (2003) 599.\\
47) G. Efstathiou: Mon. Not. Roy. Astron. Soc. \textbf{343} (2003)
L95.\\
48) J. P. Luminet: Nature \textbf{425} (2003) 593.\\
49) M. Tegmark, M. A. Strauss, M. R. Blanton, K. Abazajian, S.
Dodelson, H. Sandvik, X. Wang, D. H. Weinberg, I. Zehavi, N. A.
Bahcall, F. Hoyle, D. Schlege, R. Scoccimarro, M. S. Vogeley, A.
Berlind, T. Budavari, A. Connolly, D. J. Eisenstein, D. Finkbeiner3,
J. A. Frieman, J. E. Gunn, L. Hui, B. Jain, D. Johnston, S. Kent, H.
Lin, R. Nakajima, R. C. Nicho, J. P. Ostriker, A. Pope, R. Scranton,
U. Seljak, R. K. Sheth, A. Stebbins, A. S. Szalay, I. Szapudi, Y.
Xu, J. Annis, J. Brinkmann, S. Burles, F. J. Castander, I. Csabai,
J. Loveday, M. Doi, M. Fukugita, B. Gillespie, G. Hennessy, D. W.
Hogg, Z. Ivezic, G. R. Knapp, D. Q. Lamb, B. C. Lee, R. H. Lupton,
T. A. McKay, P. Kunszt, J. A. Munn, L. OConnell, J. Peoples, J. R.
Pier, M. Richmond, C. Rockosi, D. P. Schneider, C. Stoughton, D. L.
Tucker, D. E. Vanden Berk, B. Yanny, D. G. York: Phys. Rev. D
\textbf{69} (2004) 103501.\\
50) Y. Gong, B. Wang and Zhang, Y.Z.: Phys. Rev. D \textbf{72}
(2005) 043510.\\
51) U. Seljak, A. Slosar and P. McDonald: JCAP \textbf{10} (2006)
014.\\
52) D. N. Spergel, R. Bean, O. Dore, M. R. Nolta, C. L. Bennett, J.
Dunkley, G. Hinshaw, N. Jarosik, E. Komatsu, L. Page, H. V. Peiris,
L. Verde, M. Halpern, R. S. Hill, A. Kogut, M. Limon, S. S. Meyer,
N. Odegard, G. S. Tucker, J. L. Weiland, E. Wollack, E. L. Wright:
Astrophys. J. Suppl. \textbf{170} (2007) 377.\\
53) Q. G. Huang and M. Li: JCAP \textbf{08} (2004) 013.\\
54) C. L. Bennett, M. Halpern, G. Hinshaw, N. Jarosik, A. Kogut, M.
Limon, S. S. Meyer, L. Page, D. N. Spergel, G. S. Tucker, E.
Wollack, E. L. Wright, C. Barnes, M. R. Greason, R. S. Hill, E.
Komatsu, M. R. Nolta, N. Odegard, H. V. Peiris, L. Verde, J. L.
Weiland: Astrophys. J. Suppl. \textbf{148} (2003) 1.\\
55) K. Ichikawa, M. Kawasaki, T. Sekiguchi and T. Takahashi: JCAP
\textbf{06} (2006) 005.\\
56) E. Komatsu, J. Dunkley, M. R. Nolta, C. L. Bennett, B. Gold, G.
Hinshaw, N. Jarosik, D. Larson, M. Limon, L. Page, D. N. Spergel, M.
Halpern, R. S. Hill, A. Kogut, S. S. Meyer, G. S. Tucker, J. L.
Weiland, E. Wollack and E. L. Wright: Astrophys. J. Suppl.
\textbf{180} (2009) 330.\\
57) E. Komatsu, K. M. Smith, J. Dunkley, C. L. Bennett, B. Gold, G.
Hinshaw, N. Jarosik, D. Larson, M. R. Nolta, L. Page, D. N. Sperge,
M. Halpern, R. S. Hill, A. Kogut, M. Limon, S. S. Meyer, N. Odegard,
G. S. Tucker, J. L. Weiland, E. Wollack, and E. L. Wright:
Astrophys. J. Suppl. \textbf{192} (2011) 18.\\
58) G. Hinshaw, D. Larson, E. Komatsu, D. N. Spergel, C. L. Bennett,
J. Dunkley, M. R. Nolta, M. Halpern, R. S. Hill, N. Odegard, L.
Page, K. M. Smith, J. L. Weiland, B. Gold, N. Jarosik, A. Kogut, M.
Limon, S. S. Meyer, G. S. Tucker, E. Wollack, E. L. Wright:
\textit{Nine-Year Wilkinson Microwave Anisotropy Probe (WMAP)
Observations: Cosmological Parameter Results}, arXiv: 1212.5226.
\end{document}